\documentstyle[twoside,fleqn,espcrc2,epsf]{article}

\newcommand{\AmS}{{\protect\the\textfont2
  A\kern-.1667em\lower.5ex\hbox{M}\kern-.125emS}}

\title{Numerical Computations in the Worldsheet
       Formulation}

\author{H. Fort \\
        {Instituto de F\'{\i}sica, Facultad de Ciencias,  
        Tristan Narvaja 1674, 11200 Montevideo, Uruguay}%
        \thanks{Supported in part by CONICYT, Project No. 318.}}

\begin{document}

\begin{abstract}
The worldsheet formulation of lattice gauge theories has two appealing
features: the gauge non-redundancy and the geometrical transparency. Both
properties are profitable in order to perform numerical computations.
In the case of dynamical fermions this description offers additional 
advantages. For instance, it does not suffer from the species doubling 
problem and it involves fewer degrees of freedom.
\end{abstract}

\maketitle

\section{The Worldsheet Formulation Approach}

   Simulating  QCD
with dynamical fermions or {\em full} QCD 
is still too expensive in computer time.
Year after year the conclusion at lattice meetings is the same:
either more machine power and/or a
real improvement in algorithms
are needed to produce reliable estimates of hadron 
spectroscopy \cite{lat}.
We propose to explore a new numerical approach based
on the socalled {\em P-representation} \cite{fg}.
Staggered fermions can be
naturally introduced by considering open paths representing
string-like ``electromeson" configurations in addition
to the closed ones or loops representing pure electric flux
excitations. This gives rise to the
space of paths P of the above mentioned 
P-representation \cite{fg}.  
All the physical operators have a geometrically transparent
realization in this space.

The worldsheet formulation of lattice gauge theories  
(in terms of variables attached to the worldsheets of P-paths)
was introduced for several models \cite{abf}-\cite{afg2}.
U(1) theory \cite{abf} and 
scalar electrodynamics (SQED$_c$) \cite{abfs} were
simulated using this representation
and the results were quite encouraging.
More recently, a worldsheet description 
of gauge theories with dynamical fermions was proposed \cite{afg}.

The worldsheet partition function $Z_P$ of QED is expressed as a
sum over the worldsheets of strings or paths of the P-representation. 
That is, surfaces such that: {\bf (I)} their borders 
${\cal F}^c$ are self-avoiding polymer-like loops and {\bf (II)}
when intersected with a 
time $t=\mbox{constant}$ plane they produce
paths begining at even sites and ending at odd ones. 
This description differs from that one
gets by integrating the fields in the Kogut-Susskind action
in two features. In first place, the isolated links traversed 
in both opposite directions or ``null" links 
were ruled out and with them the myriad of different configurations
corresponding to a given configuration of worldsheets.
In second place, by virtue of the constraint {\bf (II)},
the surfaces when intersected with a 
time $t=\mbox{constant}$ plane produce only
paths begining at even sites and ending at odd ones
instead of the paths of ends with whatever parity.
This two differences translate respectively in: computer
time economy and a cure of the species doubling problem of
the Kogut-Susskind action.  

The expression of $Z_P$ is as follows \cite{foot1}:
\begin{equation}
Z_P= 
\sum_{S_{{\cal F}^c}} \, 
\sigma({\cal F}^c) \, \exp\{
-\frac{1}{2\beta} \sum_{p \in S_{{\cal F}^c}} n_p^2
\},
\label{eq:ZP0}
\end{equation}
where the sign $\sigma_{\cal F}$ is given in terms of purely 
geometric quantities of the fermionic loops ${\cal F}^c$ \cite{afg}.
In ref.\cite{afg} it was proved that $Z_P$ leads to 
the QED Hamiltonian using the transfer matrix procedure. 

Our aim was to test the worldsheet formulation of
ref. \cite{afg} as a method to simulate dynamical fermions. 
Hence, we chose the simplest lattice 
gauge theory with fermions, the Schwinger model or (1+1) QED, and
we performed a Monte Carlo simulation.  This massless model
can be exactly solved in the continuum.
However, it is rich enough to share many features 
with 4-dimensional QCD as 
the confinnement and
the breaking of the chiral symmetry with an axial anomaly.
For this reason it has been extensively used as a laboratory to 
analyze the previous phenomena and also 
to test different techniques to simulate
theories with fermions.

\section{The Schwinger Model in the Worldsheet formulation}

In $D=2$, it turns out that $\sigma({\cal F}^c)$
can be expressed in terms of the number of connected
parts $N_{{\cal F}^c}$, the lenght $L_{{\cal F}^c}$ 
and the area $A_{{\cal F}^c}$ of ${\cal F}^c$ as
$\sigma({\cal F}^c)=(-1)^{N_{{\cal F}^c}-
\frac{L_{{\cal F}^c}}{2}+A_{{\cal F}^c}}$ and all the 
non-vanishing contributions have $\sigma_{\cal F}=+1$.
The reason is that $N_{{\cal F}^c}-
\frac{L_{{\cal F}^c}}{2}+A_{{\cal F}^c}=I_{\cal F}^c$,
the number of enclosed sites by the fermionic 
loops ${\cal F}^c$ which is always even by virtue of
the above constraints {\bf (I)} and {\bf (II)} 
( see ref.\cite{afg} for more details), 
so that we omit this factor.
The fermionic paths ${\cal F}^c$ can be expressed in terms
of integer 1-forms
--attached to the links-- $f$ with three possible values: 0 and
$\pm$ 1 with the constraint that they are 
non self-crossing and closed ($\partial f = 0$) as
\begin{equation}
Z_P^{Schwinger}= 
\sum_{n} \, \sum_{f} \, 
\exp \{
-\frac{1}{2\beta} \|n\|^2
\}
\delta (f-\partial n).
\label{eq:ZP}
\end{equation}
The lattice chiral condensate per-lattice-site is defined 
as $<\bar{\chi}\chi >
=\frac{1}{2N_s}\sum_x (-1)^{x_1}
<[\hat{\chi}^\dagger 
(\mbox{x}),\hat{\chi}(\mbox{ x})]>$, where 
$N_s$ is the number of lattice sites.
The corresponding operator is realized in the 
P-representation and thus we get for 
the chiral condensate \cite{fg}:
\begin{equation}
<\bar{\chi}\chi >=
\frac{1}{2}-\frac{2{\cal N}_P}{N_s},
\label{eq:chir-cond2}
\end{equation}
where ${\cal N}_P$ is the number of connected paths at a given
time $t$.
Thus, equation (\ref{eq:chir-cond2}) allows to calculate directly
the chiral condensate simply by counting the number of 
``electromesons" we have when we intersect their world sheets
with each time slice $t$. 

\section{Monte Carlo Simulation and Results}

To generate the worldsheets of the string-like configurations
we use a Metropolis-type updating algorithm
with the Boltzman weight  
proportional to $\exp \{-\frac{1}{2\beta} \sum_p n_p^2\}$.
We simulate the model on $L\times L$ square lattices with periodic 
boundary conditions ($pbc$).

All the plaquettes $p$ belonging
to an open surface, by virtue of the self-avoiding constraint on
their borders, must have 
all their plaquettes fulfilling
the condition that the difference between integers $n_p$
of contiguous plaquettes has 3 possibilities: $\Delta n_p=\pm 1$
or 0. There are also closed surfaces
corresponding to the world sheets of loop-like pure
electric flux lines.
In two dimensions the only possibility
for a pure electric (closed) surface is to cover the entire
lattice and then being closed by the $pbc$.
Thus, all the plaquettes $p$ belonging to it
have assigned the same integer value of
$n_p = \pm 1,\pm 2,....$.
Our algorithm works as follows: whenever
there are different possible values for the integer $n_p$ 
attached to plaquette $p$ -- such that the produced surface 
fulfills the constraints {\bf (I)} and {\bf (II)}
(self avoiding frontiers and odd length string respectively) --
the present value is changed 
to one of these values and the Metropolis test is performed.
Sometimes there is only one possible value of $n_p$
consistent with the constraints
{\bf (I)} and {\bf (II)}, in such cases the $n_p$ 
is updated to it and the Metropolis
test is skipped.
Typically, 100000 sweeps per point were performed.

To compute $<\bar{\chi}\chi >$
using (\ref{eq:chir-cond2}) we have to count 
${\cal N}_P$ at each one of the $L$ time slices.
Therefore for each lattice sweep we collect $L$ 
values of $<\bar{\chi}\chi >$.

First, we checked that we get the right strong
coupling behavior both for the 
ground-state energy density $\omega_0$ and the chiral condensate 
per-lattice-site $<\bar{\chi}\chi >$. The agreement with the series 
expansion is pretty well up to $\beta=0.12$ as one can see from FIG.1.

\begin{figure}[htb]
\epsfysize=1.9 in
\epsfbox[25 185 585 675] {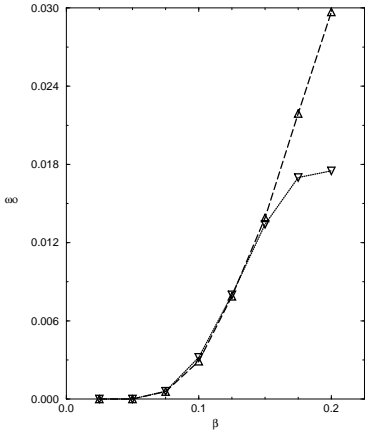}
\caption{(a) The strong coupling behavior of
$\omega_o$ on a 4x4 lattice (up triangles). 
(b) The corresponding strong coupling series (down triangles).}
\label{fig1}
\vspace{-2mm}
\end{figure}

On the other hand, the continuum theory is reached 
at zero lattice coupling, in the same way as four 
dimensional asymptotically
free theories like QCD.
It is known exactly that the chiral condensate in the 
continuum is given by   
\begin{equation}
\frac{<\bar{\psi} \psi>}{e} = \frac{e^\gamma}{2\pi^{3/2}}=0.15995.
\label{eq:chiral-cont}
\end{equation}

As long as the lattice size $L$ is increased,
the convergence of $q=<\bar{\chi}\chi >\!\!/e$ to 
its known continuum value (\ref{eq:chiral-cont}) improves. 
A property of the action of (\ref{eq:ZP}) is that 
$q$ becomes stabilized at its continuum
value well inside the weak coupling. Before reaching this
regime, this observable oscillates strongly with $\beta$. 
In FIG.2 we plot $q$ vs. $\beta$ 
for lattices of 16$\times$16, 20$\times$20, 24$\times$24
and 28$\times$28.
The value at which the simulation stabilizes 
at the weak coupling regime
decreases from around 0.29 for $L=16$ to 0.17 for $L=28$. 

\begin{figure}[htb]
\vspace{-9pt}
\epsfysize=2in
\epsfbox[25 125 395 500] {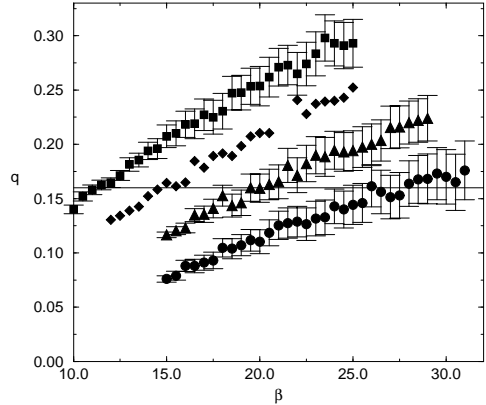}
\caption{$q=<\bar{\chi}\chi >\!\!/e$
for $L=$16 (squares), $L=20$ (diamonds), $L=24$ (triangles) 
and $L=28$ (circles). The horizontal line
correspond to the continuum limit value $\simeq 0.16$.}
\label{fig2}
\vspace{-2mm}
\end{figure}

\section{Conclusions and Final Remarks}

Our proposal was to show that the recently introduced
worldsheet formulation was a valuable and alternative tool
in order to do numerical
computations with dynamical fermions. 
The method present the following advantages: {\bf 1)} The
easiness of computation.   
{\bf 2)} Economy I: no gauge redundancy. 
{\bf 3} Economy II: it involves 
fewer degrees of freedom that the Kogut-Susskind action.
{\bf 4) } As a consequence of the constraint {\bf (II)}, 
the worldsheet action does not suffer from  
the species doubling problem of the Kogut-Susskind action.

The values for the chiral condensate are consistent 
with the exact results for the continuum theory 
(recovered at the weak coupling fixed point).
In addition, our strong coupling results 
are in agreement with the strong coupling expansion.

\end{document}